\documentclass[10pt,a4paper]{article}
\pdfoutput=1
\usepackage{graphicx}
\usepackage{amsmath}  
\usepackage{authblk}
\usepackage{hyperref}
\usepackage{fancyhdr}

\title{Shashlik calorimeters: novel compact prototypes for the ENUBET experiment}

\author{M.~Pari}
\affil{\footnotesize Universit\`a degli Studi di Padova, Via Marzolo 8, Padova, Italy\\
                     INFN — Sezione di Padova, Via Marzolo 8, Padova, Italy\\
                     CERN, Route de Meyrin 385, Geneva, Switzerland}

\author{On behalf of the ENUBET Collaboration:\\G.~Ballerini}
\author{A.~Berra} 
\author{R.~Boanta}
\author{M.~Bonesini} 
\author{C.~Brizzolari} 
\author{G.~Brunetti} 
\author{M. Calviani}
\author{S. Carturan}
\author{M.G.~Catanesi}
\author{ S.~Cecchini}
\author{A. Coffani}
\author{F.~Cindolo} 
\author{G.~Collazuol} 
\author{E.~Conti} 
\author{F.~Dal Corso} 
\author{G.~De Rosa} 
\author{C.~Delogu}
\author{A.~Gola}
\author{R. A. Intonti}
\author{C.~Jollet}
\author{Y.~Kudenko}
\author{M.~Laveder}
\author{A.~Longhin}
\author{P.F. Loverre}
\author{L.~Ludovici}  
\author{L.~Magaletti} 
\author{G.~Mandrioli}
\author{A.~Margotti}  
\author{V.~Mascagna}
\author{N.~Mauri}
\author{A.~Meregaglia}
\author{M.~Mezzetto}
\author{M. Nessi}
\author{A.~Paoloni}
\author{E.~Parozzi}
\author{L.~Pasqualini}
\author{G.~Paternoster}
\author{L.~Patrizii}
\author{C.~Piemonte}
\author{M.~Pozzato}
\author{F.~Pupilli}
\author{M.~Prest}
\author{E.~Radicioni} 
\author{C.~Riccio}
\author{A.C.~Ruggeri}
\author{G.~Sirri}
\author{M.~Soldani} 
\author{M.~Tenti} 
\author{M. Torti} 
\author{F.~Terranova}
\author{E.~Vallazza}
\author{M. Vesco}
\author{L.~Votano}

\date{\small PRESENTED AT\\\small PM2018 - 14$^\text{th}$ Pisa Meeting on Advanced Detectors\\ \small from 27$^{\text{th}}$ May to 2$^{\text{nd}}$ June 2018, La Biodola, Isola D'Elba (Italy)}

\begin{document}

\maketitle

\begin{abstract}
We summarize in this paper the detector R\&D performed in the framework of the ERC ENUBET Project. We discuss in particular the latest results on longitudinally segmented shashlik calorimeters and the first HEP application of polysiloxane-based scintillators.
\end{abstract}

\section{Introduction}
The ENUBET project~\cite{art:aNovel,art:SPSC} aims to develop a neutrino source based on tagging
of large angle positrons from $K_{e3}$ decays (i.e. $K^+ \rightarrow e^+\,\pi^0\,\nu_e$) in an instrumented decay tunnel.
This would lead to a reduction of the systematic uncertainties on the knowledge of the initial neutrino flux to $\sim 1$\%,
i.e. one order of magnitude lower than present neutrino beams.
This facility would highly impact on accelerator neutrino physics, allowing for an unprecedented precision
on the $\nu_e$ and $\bar{\nu}_e$ cross section measurement. 
The decay tunnel to be implemented consists in an hollow cylinder with a length of $40$~m and a 1 m radius.
Particle transport and interactions in the beamline were simulated with G4beamline~\cite{art:g4beamline}
while the interactions in the instrumented decay tunnel and the detector response were simulated with GEANT4.
The expected rate at the detector (positron tagger) is $200$~kHz$/$cm$^2$ and $e/\pi$ separation at the $<$3\% level is needed to reject the
pion background due to beam halo and to other decay modes of kaons. This separation is achieved by means of longitudinally segmented calorimeters.
These requirements constrain the
detector technology, which must be based on radiation hard components with $O(10~\text{ns})$ recovery time
and a $\sim10$~cm$^2$ granularity.
Cost effectiveness is a key parameter since the number of modules for the instrumentation of the tunnel is $10^5$.

\section{Compact shashlik calorimeters}

Shashlik type calorimeters~\cite{art:shashOrigin}, namely scintillator-based sampling calorimeters readout by
optical wavelength shifter (WLS) fibers crossing the iron-scintillator tiles, fulfill the cost effectiveness and fast response constraints mentioned above.
A pioneering compact solution has been developed in the context of the SCENTT INFN R\&D~\cite{art:aCompact,art:berraShash}, leading to
the baseline ENUBET single module prototype, called Ultra Compact Module (UCM).
\begin{figure}
\centering
\includegraphics[width=0.9\linewidth]{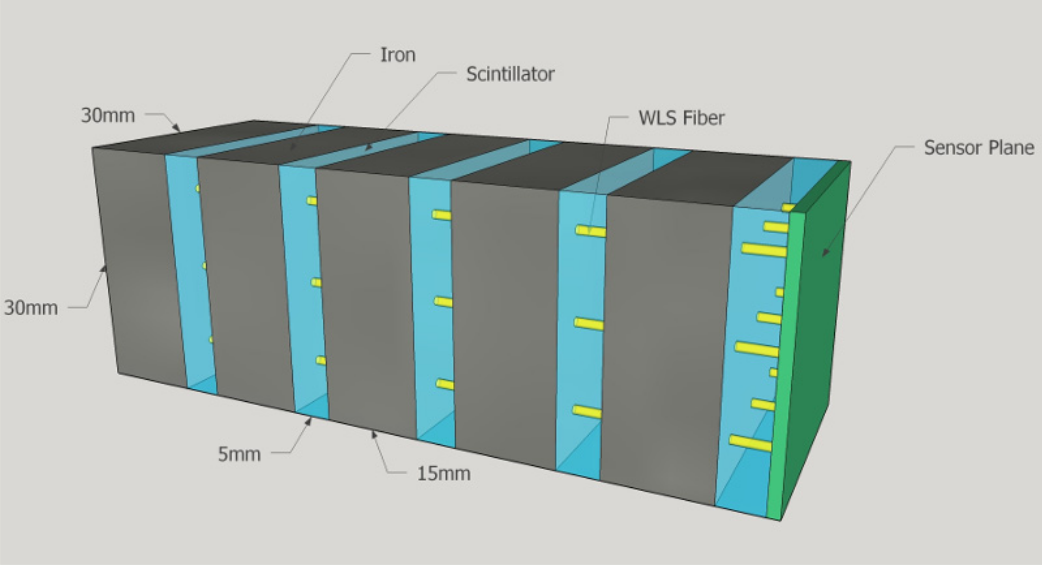}
\caption{Schematics of a single detector module (UCM: Ultra Compact Module) for ENUBET.}
\label{fig:baseline}
\end{figure}
Fig.~\ref{fig:baseline} shows a schematic of the baseline prototype for the ENUBET experiment, composed by $5$, $15$~mm thick, iron layers interleaved by $5$~mm thick
plastic scintillator tiles. The length of the module is $10$~cm, corresponding to $4.3$ radiation lengths ($X_0$). The UCM has a transverse area of $3\times3$~cm$^2$,
which corresponds to $1.7$ Moliere radii. A SiPM-based readout system was integrated directly in the bulk of
the detector: every WLS fiber ($9$ in total) is readout by one single SiPM. The ENUBET SiPMs are produced by FBK with a $1$~mm$^2$ sensitive area and $2500$ $20\times20~\mu$m cells.

\section{Beam tests at CERN-PS T9}

A positron tagger prototype has been tested assembling $56$ UCMs
in a $7\times4\times2$ structure: in the longitudinal direction $7$ UCMs sample the development of the electromagnetic and hadronic showers.
The calorimeter ($30.1~X_0$ and $3.09$ interaction lengths)
was tested at CERN-PS T$9$ beamline in November 2016~\cite{art:novTB}.
By selecting electrons through an upstream Cherenkov counter, the resolution of the calorimeter was measured in the $1$-$5$~GeV energy range,
i.e. in the range of interest for neutrino physics applications.
The results for electrons ($(15.7\pm0.7)\%/\sqrt{\text{GeV}}$) and the pattern of energy deposition of pions are in good agreement
with the Monte Carlo simulation
and confirm the detector specifications of ENUBET.
The resolution and $e/\pi$ discrimination were tested for particles impinging with a tilt angle from $0$ to $250$~mrad reproducing the
conditions envisaged in the ENUBET decay tunnel.

\section{Test of SiPM radiation hardness}

Since the SiPMs are embedded in the bulk of the UCM, it is crucial to
verify that radiation effects do not compromise the performance of the ENUBET calorimeter. Nuclear Counter Effects
has been demonstrated to be negligible in~\cite{art:berraShash}. During the lifetime of the experiment the SiPM
will integrate a neutron fluence of $O(10^{11})$~$1$~MeV-equivalent neutrons$/$cm$^2$.
SiPMs are particularly sensitive to neutron
damage and a dedicated irradiation campaign at the CN facility at INFN-LNL (Legnaro) was performed in June 2017.
Neutron fluences up to $10^{12}$~n$/$cm$^2$ were integrated by impinging $5$~MeV protons from the CN Van de Graaff generator to a thick Beryllium target.
From the photodetection characterization of the irradiated SiPMs, we demonstrated that the sensitivity to single photon is lost after an integrated dose
of $3\times10^9$~n/cm$^2$ (1 MeV equivalent).
UCM prototypes read out by the neutron irradiated SiPM boards were tested at CERN in October 2017 (T9 beamline). The test demonstrated
that  the calorimetric performance of the detector are not compromised
by irradiation in the $O(10^{11})$ n/cm$^2$ regime. 
The energy response of the UCMs to a beam of $\pi^-$, $e^-$ and $\mu^-$ shows that the ratio between the MIP peak and electron peak remains constant.
In particular it is still possible to distinguish the MIP peak from the pedestal after the irradiation.
The overall gain reduction of the irradiated boards can be recovered by increasing the overvoltage within the SiPM operational range.

\section{Polysiloxane scintillators}

As an alternative to standard plastic scintillator, a polysiloxane-based
scintillator~\cite{art:carturan,art:quaranta} was employed in the
construction of the UCMs. Polysiloxane doped with scintillating material
is liquid at high temperature and reaches a soft solid state at room temperature.
Polysiloxane-based
scintillators ease in a substantial manner the construction of shashlik calorimeters because they can be poured over the fibers without drilling the scintillator.
We demonstrated that this system has the same optical coupling to fibers than injection molded plastic scintillators. 
This material has a radiation hardness ten times larger than standard plastic scintillators and remains transparent up to $10$~kGy.
It has however a light yield that is $30$\% the yield of the best Polyvinyltoluene-based scintillators (e.g. EJ-200).
A $12$~UCMs prototype, assembled in a $3\times2\times2$ structure, has been tested during the October 2017 test beam
at T$9$. The polysiloxane-based UCMs had a thickness of $15$~mm, i.e. three times the thickness of the baseline prototype to compensate for the reduced light yield of polysiloxane.
The energy resolution in the few GeV range is $16\%/\sqrt{\text{GeV}}$ and it is comparable with the prototype based on plastic scintillators.

\section*{Acknowledgments}
This project has received funding from the European Union’s Horizon $2020$ Research and Innovation 
programme under Grant Agreement no. $654168$ and no. $681647$.
 

\end{document}